\documentclass[12pt]{iopart}
\usepackage{graphicx, bm, xcolor, amssymb}

\begin{document}

\title{Non-adiabatic dynamics of the entanglement entropy in a symmetry-breaking Haldane insulator}
\author{Junjun Xu}
\address{Institute of Theoretical Physics, University of Science and Technology Beijing, Beijing 100083, China}
\ead{\mailto{jxu@ustb.edu.cn}}
\date{\today}

\begin{abstract}
We study the non-adiabatic dynamics of a typical symmetry-protected topological phase-the Haldane insulator phase with broken bond-centered inversion. By continuously breaking the middle chain, we find the gap closes at a critical point in the deep Haldane insulator regime with a change of particle number partition of the left or right system. The adiabatic evolution fails at this critical point and we show how to predict the dynamics of the entanglement entropy near this point using a two-level model. These results show that one can find a critical regime where the entanglement measurement is relatively robust against perturbation that breaks the protecting symmetries in the Haldane insulator. This is in contrast to the common belief that the symmetry-protected topological phases are fragile without the protecting symmetries.
\end{abstract}

%\pacs{}
\maketitle

\section{Introduction}
The robustness of symmetry-protected topological (SPT) phases to symmetry-respecting perturbations makes them promising candidates for quantum computing~\cite{Chiu}. Recently, the topological characteristics of these phases have been extended to out-of-equilibrium regime~\cite{Cooper}, and more recent studies show that their topologies might be robust even when the initial state breaks the protecting symmetries~\cite{Marks}. Thus it would be of both theoretical and technical interests to understand to what extent these phases are robust during dynamical evolution.

In this paper, we try to focus on this question: how robust the entanglement measurement is if we not only break the protecting symmetries of the initial state, but also the time-dependent Hamiltonian? Naively, one might imagine the SPT state will adiabatically evolve to a trivial state which eliminates the entanglement. However, we will show that one might find a critical point where this adiabatic evolution fails, thus leads to relatively robust entanglement measurement.

We consider the Haldane phase in a more controllable cold atoms system, which was firstly considered by Dalla Torre {\it et al.}, termed as the Haldane insulator (HI), and the phase diagram has been well studied~\cite{Dalla, Rossini}. This HI phase shows non-trivial properties similar to the Haldane phase in spin-1 antiferromagnetic Heisenberg chains~\cite{Haldane, Affleck} (the non-local string order, a two-fold degeneracy of the entanglement spectrum, etc.), but with a revised protecting bond-centered (BC) inversion symmetry~\cite{Berg, Li, Pollmann, Deng, Ejima}. Recent theoretical progress has shed light on realizing this bosonic HI phase by providing strong enough long-range interaction using Feshbach resonance~\cite{Xu}. 

In the following, we will consider the middle chain-breaking dynamics of the HI phase with broken BC inversion. This dynamical process has been considered by Pollmann {\it et al.} in the spin-1 chain respecting the BC inversion symmetry, and they find a lower bound ($\log2$) for the half-system von Neumann entanglement entropy by adiabatically breaking the middle chain. This entanglement measurement is due to the double degeneracy of the entanglement spectrum, and is robust with symmetries even not protecting the edge modes and string order~\cite{Pollmann}. In this work, we will focus on the entanglement entropy, and show to what extent this measurement is robust. To our knowledge, this problem has not been fully understood before.

This paper is organized as follows. First, we introduce the middle chain-breaking experiment in an extended Bose-Hubbard system and consider its adiabatic properties. We show the broken system has a vanishing energy gap in the deep HI phase regime, as a result of the degeneracy of two trivial product states appear in this phase.  Then we consider the non-adiabatic dynamics. We get the full numerical evolution of the entanglement entropy by integrating the time-dependent Schr\"odinger equation. We find the entanglement entropy is relatively robust in the deep HI regime. We further give an analytical prediction to the entanglement entropy in this regime by mapping the system to a two-level model. At last, we give our conclusions and mention the relevant experiment measurements.

\section{The middle-chain breaking and its adiabatic properties}
We consider a chain-breaking experiment on the extended Bose-Hubbard model $H=H_0+\gamma H'$ with
\begin{eqnarray}
H_0=&&-J\sum_{i\neq L/2}(b^\dagger_{i}b_{i+1}+\mathrm{H.c.})+\frac{U}{2}\sum_i n_i(n_i-1)\nonumber\\
&&+V\sum_{i\neq L/2}n_{i}n_{i+1},\nonumber\\
H'=&&-J(b^\dagger_{L/2}b_{L/2+1}+\mathrm{H.c.})+Vn_{L/2}n_{L/2+1}
\label{eq:eq1}
\end{eqnarray}
describing the broken and linking Hamiltonian. We consider a chain length of $L$. Here $J$ characterizes the nearest-neighbor hopping, and $U$, $V$ are the on-site and nearest-neighbor interaction strength, with $n_i=b_i^\dagger b_i$ the bosonic particle number operator at site $i$. The parameter $\gamma\in[0,1]$ characterizes the strength of middle bond. At $\gamma=1$, we recover the conventional extended Bose-Hubbard model. At $\gamma=0$ this chain breaks into two length-$L/2$ subsystems.

In this work, we run an exact diagonalization calculation with chain length $L=12$. This is a typical system size can be realized in current cold atoms experiments~\cite{Rispoli}. To observe the HI phase, we constrain the maximum on-site particle number to 2, the state space of which can then be mapped to an effective spin-1 system. Experimentally, this constraint can be achieved by appropriately include the Feshbach resonances in this system as recently proposed in~\cite{Xu}. We consider a half-filling case with total particle number $N=L$, which is equivalent to the total spin sector $S_{tot}^z=0$ in the mapped spin system.  We fix the edge particle number to be $2$ and $0$ in the following calculation, which helps to break the ground state degeneracy in the HI and density wave phases, and reducing the particle-hole excitations at the edge. In experiment this can be done by reducing or increasing the local potential at the left or right edge of the optical lattice. Such edge configuration has also been considered in previous studies, and the HI is found sandwiched between Mott insulator and density wave phases (for on-site interaction $U/J=4.0$ the HI phase appears in the range $2.1\lesssim V/J\lesssim 3.0$)~\cite{Dalla, Rossini, Ejima}.

\begin{figure}[t]
 \begin{center}
 \includegraphics[width=0.6\textwidth]{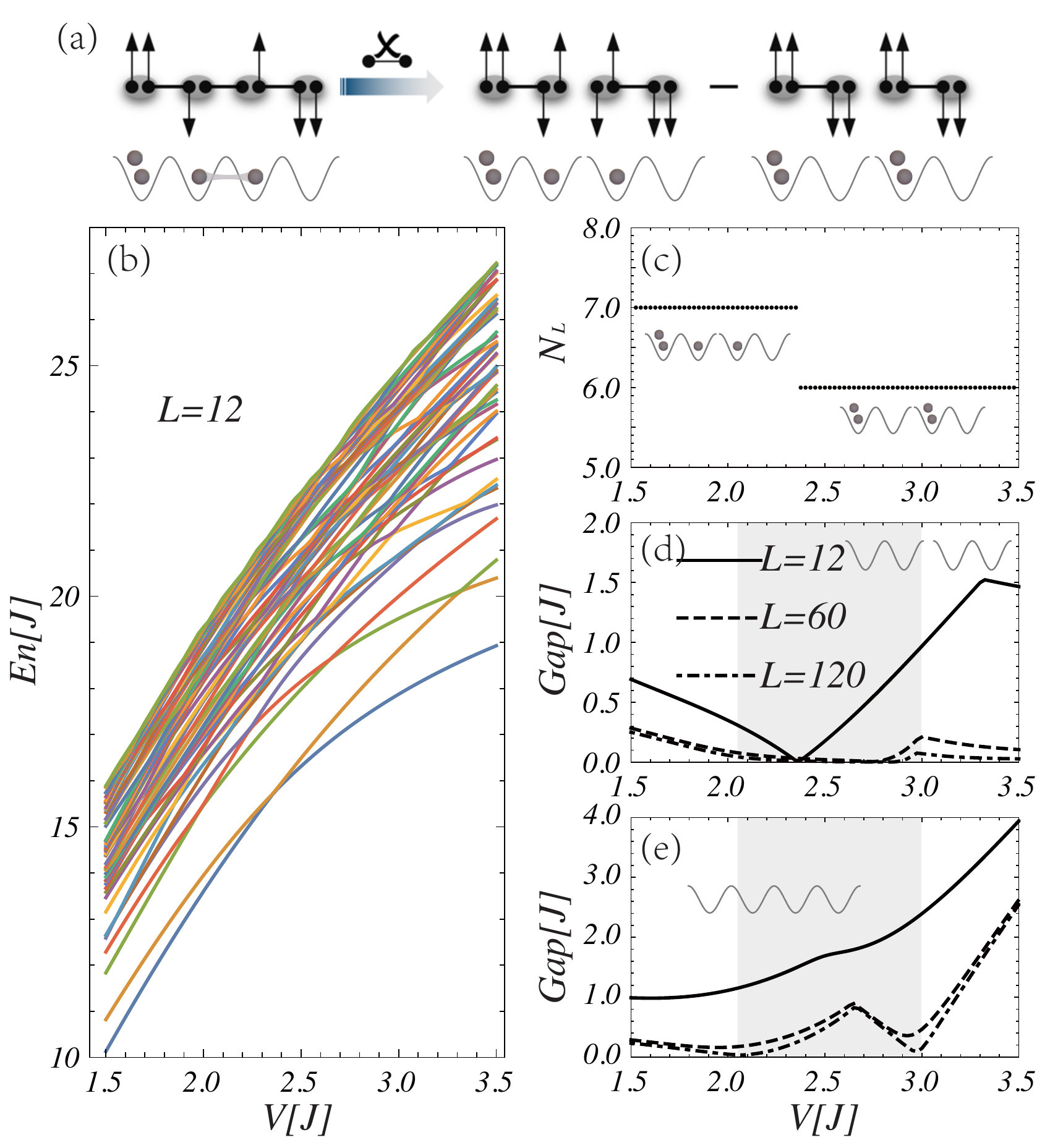}
  \caption{Chain-breaking experiment of an open Haldane insulator. (a) An Affleck-Kennedy-Lieb-Tasaki (AKLT) state description for middle chain-breaking, where the black dots represent effective spin-1/2 particles. By adiabatically breaking the middle bond, the AKLT state evolves to one of the trivial states as the ground state. (b) The energy spectrum after breaking the middle chain for different nearest-neighbor interaction strength $V$. The lowest level crossing happens at around $V/J\approx 2.4$. (c) Ground state particle number of the left-half system $N_L$ after breaking the middle chain for different $V$. The particle number shifts at the critical point $V/J\approx 2.4$, which is within the Haldane insulator regime (shaded region in d and e). The energy gap after and before breaking the middle chain is shown in (d) and (e). The results are obtained for $U/J=4$ with the chain length $L=60/120$ based on DMRG calculations using ALPS package~\cite{White, Schollwock, alps}.}
  \label{fig:fig1}
 \end{center}
\end{figure}

We note that even though the above choice of edge configuration stabilizes the HI state, it in fact has already broken the BC inversion symmetry. Thus even though the linking Hamiltonian $H'$ conserves BC inversion, the system loses its protecting symmetry all along with the time evolution. So by breaking the middle chain, adiabatically the HI will  evolve to a trivial product state (here by ``product state" we mean the state is composed of left half-system state producting the right half-system state). However, since there are two product states competing as illustrated in Fig.~\ref{fig:fig1}a, this adiabatic evolution will not be valid if these two states are degenerate. We show the energy spectrum for the lowest $40$ eigenstates in Fig.~\ref{fig:fig1}b. We find a ground state level crossing around $V/J\approx 2.4$, which is due to the competition of these two product states. This is consistent with the particle number partition in the left-half system in Fig.~\ref{fig:fig1}c. Obviously, these two product states are adiabatically connected to the Mott insulator and density wave states respectively by changing the nearest-neighbor interaction $V/J$. In Fig.~\ref{fig:fig1}d and e we show the energy gap after and before breaking the middle chain. As the chain length increases, the level crossing extends to the full HI regime as shown in Fig.~\ref{fig:fig1}d (shaded region), with the regime coincides with Fig.~\ref{fig:fig1}e. These findings suggest that one can find regimes within the deep HI regime where the adiabatic evolution fails, and thus protect the non-trivial state even without the BC inversion symmetry. In the following, we will focus on this regime and try to understand this non-adiabatic dynamics.

\section{Non-adiabatic dynamics of the entanglement entropy}

\subsection{Full time evolution}
We consider the full time evolution in this section, to show whether the experiment is able to follow the above adiabatic processes.  We consider the dynamical breaking of middle chain which takes a linear form $\gamma(t)=1-\Gamma t$, where $\Gamma$ characterizes the rate of this breaking. The time-dependent Hamiltonian can be written as $H(t)=H_0+(1-\Gamma t)H'$. The wave function can be expanded using the orthogonal eigenstates at $t=0$,
\begin{equation}
|\psi(t)\rangle=\sum_n c_n(t)e^{-i E_n^0 t}|\psi_n^0\rangle,
\label{eq:fullansatz}
\end{equation}
where $|\psi_n^0\rangle$ and $E_n^0$ are the eigenvector and eigenvalues of the system at $t=0$, i.e., $(H_0+H')|\psi_n^0\rangle=E_n^0|\psi_n^0\rangle$. Under Eq.~(\ref{eq:fullansatz}), the Schr\"odinger equation becomes a set of coupled equations about the coefficients $c_n(t)$,
\begin{equation}
i\dot{c}_n(t)=-\Gamma t \sum_m\langle n|H'|m \rangle c_m(t)e^{-i(E_m^0-E_n^0) t}.
\end{equation}
With the initial condition $c_n(0)=\delta_{n,0}$ the above equations give the full time evolution of the system.

\begin{figure}[t]
 \begin{center}
 \includegraphics[width=0.7\textwidth]{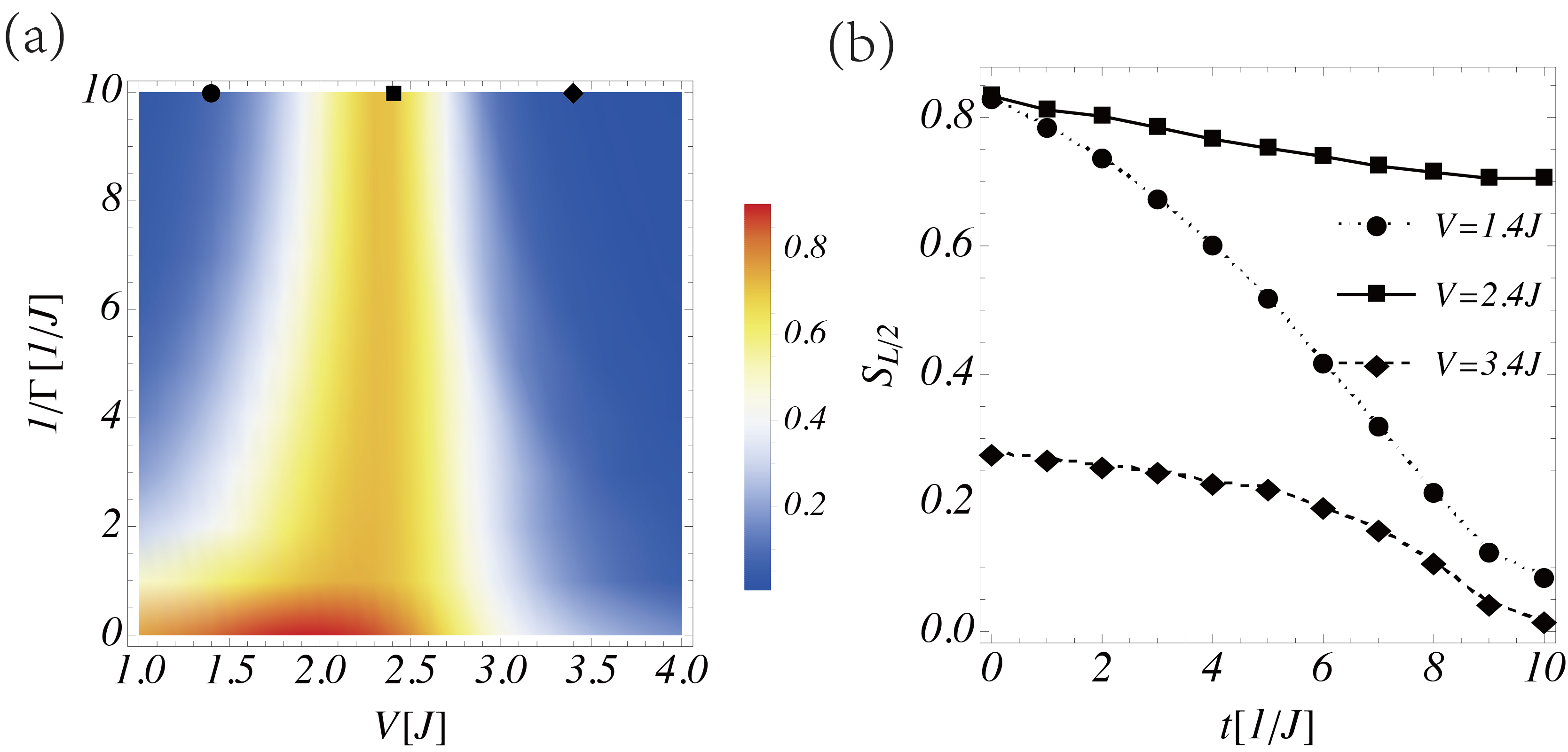}
  \caption{(a) The final-state half entanglement entropy $S_{L/2}$ at $t=1/\Gamma$ as a function of nearest-neighbour interaction $V$ and inverse breaking rate $1/\Gamma$ for on-site interaction $U/J=4$ and chain length $L=12$.  (b) Full time evolution of $S_{L/2}$ for different nearest-neighbor interaction $V/J=1.4, 2.4, 3.4$ at breaking rate $\Gamma/J=0.1$. The dark points in (a) correspond to the data in (b) at $t=10/J$. The entanglement entropy is relatively robust in the deep Haldane insulator regime at around $V/J\approx 2.4$.}
  \label{fig:fig2}
 \end{center}
\end{figure}

The more important and experimentally relevant quantity is the half-system von Neumann entanglement entropy $S_{L/2}=-\mathrm{Tr}\left[\rho_l\log\rho_l\right]$, where $\rho_l=\mathrm{Tr}_r|\psi(t)\rangle\langle\psi(t)|$ is the reduced density matrix of the left-half system with the trace over the right-half system. We show the final-state entanglement entropy $S_{L/2}$ as a function of inverse breaking rate $1/\Gamma$ and nearest-neighbor interaction $V$ in Fig.~\ref{fig:fig2}a. We pick three typical values $V/J=1.4, 2.4, 3.4$ at breaking rate $\Gamma/J=0.1$ and show the dynamics of $S_{L/2}$ in Fig.~\ref{fig:fig2}b. At the deep HI regime around $V/J\approx 2.4$, we find the entanglement entropy is relatively robust and evolves to a finite value.

\subsection{A two-level prediction}
The above full numerical calculations are time consuming and lack of physical insights. Since there are two nearly degenerate states for the broken Hamiltonian $H_0$ in the deep HI regime, we can describe the physics by a two-level model. We label these two states as $|0\rangle$ and $|1\rangle$, and the system can then be projected into these two subspaces as
\begin{eqnarray}
H_{eff}(t)=\left[\begin{array}{cc}
\langle 0|H|0\rangle& \langle 0|H|1\rangle\\
\langle 1|H |0\rangle & \langle 1|H|1\rangle
\end{array}\right].
\end{eqnarray}
We define $\langle 1|H|1\rangle-\langle 0|H|0\rangle=h-2\beta (1-\Gamma t)$, $\langle 0|H|1\rangle=\alpha (1-\Gamma t)$, where $h=\langle 1|H_0|1\rangle-\langle 0|H_0|0\rangle$ is the gap of $H_0$ between these two states. Here $\alpha=\langle 0|H'|1\rangle$ and $\beta=(\langle 0|H'|0\rangle-\langle 1|H'|1\rangle)/2$ come from the hopping and nearest-neighbor interaction terms respectively. In general, $\alpha$ could be a complex number, but we can always drop the complex phase factor into state $|0\rangle$ or $|1\rangle$ and makes $\alpha$ real. Thus in the following we choose $\alpha$ to be its modulus. The physics should not change if we shift to a rotating frame by a unitary transformation $U=e^{i\langle 0|H_0|0\rangle t+i\left(\langle 0|H'|0\rangle+\langle 1|H'|1\rangle\right)\left(t-\Gamma t^2/2\right)/2}$, and the above Hamiltonian becomes more symmetric as
\begin{eqnarray}
\tilde{H}_{eff}(t)=H_{eff}+i\dot{U}U^\dagger=\left[\begin{array}{cc}
\beta (1-\Gamma t) & \alpha (1-\Gamma t)\\
\alpha (1-\Gamma t) & -\beta (1-\Gamma t)+h
\end{array}\right].
\label{eq:twolevel}
\end{eqnarray}

Let's first look at the asymptotic behavior at $h/J\to0$ (which is true in the thermodynamic limit of the HI phase and also for the critical point in our finite system). Under this condition, the time evolution of the Hamiltonian Eq.~(\ref{eq:twolevel}) is readily solved and the solution gives
\begin{eqnarray}
\psi_+(t)=\frac{e^{i\omega(t-\Gamma t^2/2)}}{\sqrt{\alpha^2+(\beta-\omega)^2}}\left(\begin{array}{c}
\beta-\omega\\
\alpha
\end{array}\right)
\label{eq:psi0}
\end{eqnarray}
for initial state as the ground state, and
\begin{eqnarray}
\psi_-(t)=\frac{e^{-i\omega(t-\Gamma t^2/2)}}{\sqrt{\alpha^2+(\beta+\omega)^2}}\left(\begin{array}{c}
\beta+\omega\\
\alpha
\end{array}\right)
\label{eq:psi1}
\end{eqnarray}
for initial state as the first excited state. Here $\omega=\sqrt{\alpha^2+\beta^2}$. Thus it is obvious that besides a dynamical phase, the probabilities of the system in these two states $|0\rangle$ and $|1\rangle$ are constant with time. It means in the limit $h/J\to 0$, the reduced density matrix $\rho^{\pm}_l=\mathrm{Tr}_r|\psi_{\pm}(t)\rangle\langle\psi_{\pm}(t)|$ does not evolve with time and thus the half entanglement entropy of each state $\psi_{\pm}(t)$ is a constant and robust to the middle-bond breaking.

For finite $h$, we consider the long-time nearly adiabatic limit $\Gamma/J\to 0$. In this case we have a rather slow breaking rate and the dynamics is mainly determined by the details of time $t$ close to $1/\Gamma$. So the physics is determined by the region $\beta(1-\Gamma t)\ll h$, and we have the following Hamiltonian
\begin{eqnarray}
H_{ad}(t)=\left[\begin{array}{cc}
0 & \alpha (1-\Gamma t)\\
\alpha (1-\Gamma t) & h
\end{array}\right].
\label{eq:hamad}
\end{eqnarray}
To get the dynamics of $H_{ad}$, we try to expand the solution as a superposition of orthogonal states Eq.~(\ref{eq:psi0}) and (\ref{eq:psi1}) with $\beta=0$,
\begin{eqnarray}
\psi_{ad}(t)&&=c_+(t)\psi_+(t)+c_-(t)\psi_-(t) \nonumber\\
&&=c_+(t)\frac{e^{i\theta(t)}}{\sqrt{2}}
\left(\begin{array}{c}
-1\\
1
\end{array}\right)
+c_-(t)\frac{e^{-i\theta(t)}}{\sqrt{2}}
\left(\begin{array}{c}
1\\
1
\end{array}\right),
\nonumber
\end{eqnarray}
where $\theta(t)=\alpha(t-\Gamma t^2/2)$. The time-dependent Schr\"odinger Equation then becomes
\begin{eqnarray}
i\dot{c}_+e^{i\theta}-i\dot{c}_-e^{-i\theta}&&=0,\\
i\dot{c}_+e^{i\theta}+i\dot{c}_-e^{-i\theta}&&=hc_+e^{i\theta}+hc_-e^{-i\theta},
\end{eqnarray}
which can be reduced to the following second-order differential equation
\begin{equation}
\ddot{c}_++i\left[h+2\alpha(1-\Gamma t)\right]\dot{c}_+-\alpha h(1-\Gamma t)c_+=0,
\label{eq:longtime}
\end{equation}
with
\begin{eqnarray}
c_-=e^{2i\theta}\left(\frac{2i}{h}\dot{c}_+-c_+\right).
\label{eq:cm}
\end{eqnarray}
The solution of Eq.~(\ref{eq:longtime}) has an additional phase term. This can be seen by setting $t\to\pm\infty$, and Eq.~(\ref{eq:longtime}) becomes $2i\dot{c}_+-hc_+=0$ with the solution $c_+(t\to\pm\infty)=e^{-iht/2}$. This phase can be removed by defining
\begin{eqnarray}
c_+(t)=e^{-iht/2}\tilde{c}_+(t),
\label{eq:cp}
\end{eqnarray}
and Eq.~(\ref{eq:longtime}) can then be reduced to
\begin{equation}
\ddot{\tilde{c}}_++2i\alpha(1-\Gamma t)\dot{\tilde{c}}_++h^2\tilde{c}_+/4=0.
\end{equation}
By defining $z=-e^{i\pi/4}(1-\Gamma t)\sqrt{\alpha/\Gamma}$ and $\nu=h^2/(\Gamma \alpha)$, the above equation is equivalent to the Hermite differential equation
\begin{equation}
\ddot{\tilde{c}}_+(z)-2z\cdot\dot{\tilde{c}}_+(z)-i\nu/4\cdot\tilde{c}_+(z)=0,
\end{equation}
the general solution of which is readily written as a linear combination of Hermite polynomial $H_\lambda(z)$ and confluent hypergeometric function $M(\lambda_1,\lambda_2,z)$~\cite{math}
\begin{equation}
\tilde{c}_+(z)=a H_{-i\nu/8}\left(z\right)+b M\left(\frac{i\nu}{16},\frac{1}{2},z^2\right),
\end{equation}
where $a$ and $b$ are constants to be determined by the initial condition. Since we are interested in the nearly adiabatic side, we have the initial condition
\begin{eqnarray}
\left|\tilde{c}_+\left(z\left(t\to-\infty\right)\right)\right|=1, |\tilde{c}_-\left(z\left(t\to-\infty\right)\right)|=0.
\label{eq:initialcondition}
\end{eqnarray}
Consider the asymptotic behaviors $H_\lambda(z\left(t\to-\infty\right))=2^\lambda z^\lambda$, $M(\lambda_1,\lambda_2,z\left(t\to-\infty\right))=\Gamma(\lambda_2)(e^zz^{\lambda_1-\lambda_2}/\Gamma(\lambda_1)+(-z)^{-\lambda_1}/\Gamma(\lambda_2-\lambda_1))$, and plug these into the initial condition Eq.~(\ref{eq:initialcondition}) we have the coefficients
\begin{eqnarray}
a=-2^{\frac{i\nu}{8}}e^{-\frac{\pi\nu}{32}}, b=\frac{e^{-\frac{3\pi\nu}{32}}}{\sqrt{\pi}}\left(1+e^{\frac{\pi\nu}{8}}\right)\Gamma\left(\frac{1}{2}-\frac{i\nu}{16}\right).\nonumber
\end{eqnarray}
Then from Eq.~(\ref{eq:cm}) and (\ref{eq:cp}) we obtain the time-dependent solution of Hamiltonian $H_{ad}$.

\begin{figure}[t]
 \begin{center}
 \includegraphics[width=0.6\textwidth]{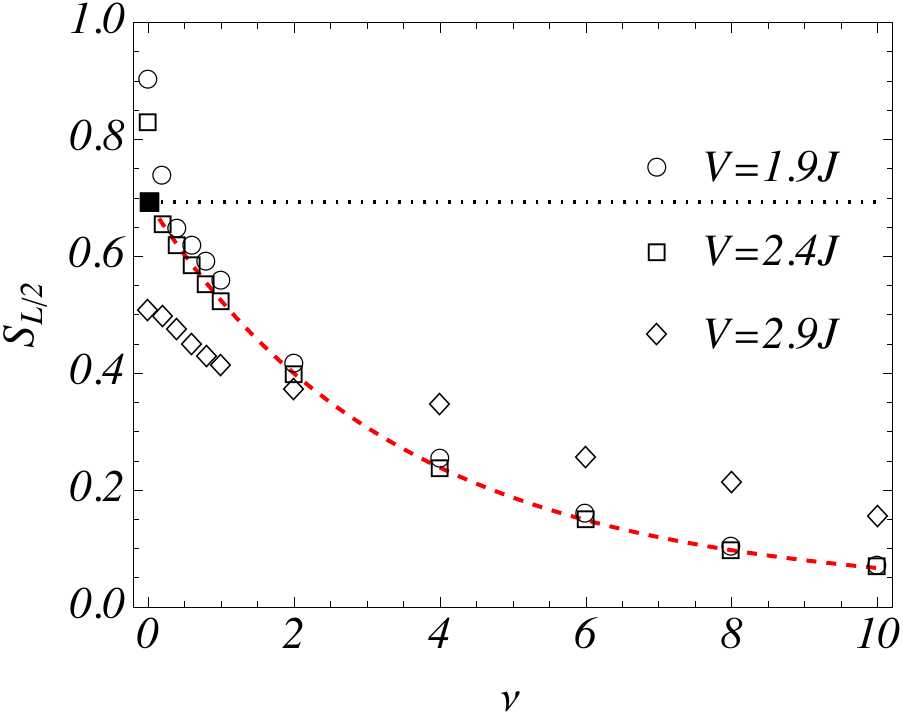}
  \caption{The half entanglement entropy $S_{L/2}$ as a function of $\nu=h^2/(\Gamma\alpha)$. The data points are from full numerical calculations and the dashed red line is our two-level prediction. The dotted black line labels $S_{L/2}=\log 2$. These results are for $U/J=4$ and chain length $L=12$. For each $V$ we first calculate the gap $h$ and coupling parameter $\alpha$, then choose different breaking rate $\Gamma$. For fixed $V$, smaller $\nu$ corresponds to larger $\Gamma$. Around the critical point $V/J\approx 2.4$ the non-adiabatic dynamics is well described by our two-level prediction with the solid square point labels the result for $\Gamma/J=0.1$ as in Fig.~\ref{fig:fig2}b.}
  \label{fig:fig3}
  \end{center}
\end{figure}

Since the two nearly degenerate states $|0\rangle$ and $|1\rangle$ at $t=1/\Gamma$ are trivial product sates and have different particle number partition for the left and right half systems, they contribute to the entanglement entropy only by their probabilities. According to the above solution, we arrive the final probability at the lowest energy state $|0\rangle$
\begin{eqnarray}
P_0&&=\left|c_+(1/\Gamma)e^{i\theta}-c_-(1/\Gamma)e^{-i\theta}\right|^2/2\nonumber\\
&&=\frac{e^{-\frac{3\pi\nu}{16}}}{2\pi}\cdot\left|\frac{e^{\frac{\pi\nu}{16}}\pi}{\Gamma(\frac{1}{2}+\frac{i\nu}{16})}+\frac{4(-1)^{3/4}e^{\frac{\pi\nu}{16}}\pi}{\sqrt{\nu}\Gamma(\frac{i\nu}{16})}-\left(1+e^{\frac{\pi\nu}{8}}\right)\Gamma\left(\frac{1}{2}-\frac{i\nu}{16}\right)\right|^{2}.\nonumber
\end{eqnarray}
Then we have the half entanglement entropy $S_{L/2}=-P_0\log P_0-(1-P_0)\log(1-P_0)$. In Fig.~\ref{fig:fig3} we show $S_{L/2}$ as a function of $\nu$ for different nearest-neighbor interaction strength $V$. Our analytical prediction is shown as the dashed red line and coincides well with the full numerical calculation at the deep HI regime around $V/J\approx2.4$, except at around $\nu=0$ where the long-time nearly adiabatic limit fails. The solid square point labels the result for sufficient slow breaking rate $\Gamma/J=0.1$ as in Fig.~\ref{fig:fig2}b. For larger $\nu$ one needs even slower breaking rate. This indicates the entanglement measurement is relatively robust in this deep HI regime.

\section{Conclusions}
We have studied the middle chain-breaking dynamics of the HI with broken BC inversion symmetry. We find a critical point within the deep HI regime, where the adiabatic evolution fails as a result of energy level crossing. We show how to understand this non-adiabatic dynamics using a simple two-level model. We give an analytical prediction to the entanglement entropy and find it is relatively robust in this regime. We note that if we rotate the state space to $\left(|0\rangle\pm|1\rangle\right)/\sqrt{2}$ in our two-level prediction, the Hamiltonian near the adiabatic limit $H_{ad}(t)$ reminds us the one of a Landau-Zener system, the infinite-long-time diabatic probability of which is readily solved~\cite{Landau, Zener}. In this paper we provide an alternative analytical solution to the full time evolution. This result is quite general, and should find applications in similar Landau-Zener-like systems.  Our results suggest the HI as an ideal system to study the non-equilibrium dynamics of SPT phases with broken symmetries.

The entanglement entropy has been considered theoretically~\cite{Abanin, Daley}, and is realized recently in cold atoms experiments, where a combination of the single-site-resolved microscope and many-body quantum interference is carried out to directly measure the second-order R\'enyi entropy $S_2=-\log\mathrm{Tr}(\rho_A^2)$~\cite{Islam, Kaufman}. This R\'enyi entropy $S_2$ provides a lower bound for the von Neumann entropy we considered here and shares similar behaviors. The generalization of our prediction to the second-order R\'enyi entropy based on the two-level model is straightforward. We also note that since the number partitions are distinct for the two lowest nearly degenerate levels in the deep HI regime, one can directly get the von Neumann entropy by measuring its number entanglement $S_n$ of half system~\cite{Lukin}.

\ack
This research is supported by Fundamental Research Funds for the Central Universities (No. FRF-TP-19-013A3).

\end{document}